\documentclass[%
 reprint,
 amsmath,amssymb,
 aps,
]{revtex4-2}

\usepackage{natbib}

\usepackage{graphicx}
\usepackage{dcolumn}
\usepackage{bm}

\usepackage{booktabs}
\usepackage{geometry}
\usepackage{amsthm}
\usepackage{mdsymbol}
 \usepackage{epstopdf}
\usepackage{float}
\usepackage{subfigure} 
\usepackage{color}
\usepackage[colorlinks,
            linkcolor=blue,
            anchorcolor=blue,
            citecolor=blue,
            urlcolor=blue]{hyperref}

\usepackage{colortbl}
\usepackage{url}
\usepackage{arydshln}
\usepackage{array}
\usepackage{appendix}
\usepackage{inputenc}

\begin{document}

\preprint{APS/123-QED}

\title{Conceptual design and science cases of a juggled interferometer for gravitational wave detection}

\author{Bin Wu$^1$, Tomohiro Ishikawa$^1$, Shoki Iwaguchi$^1$, Ryuma Shimizu$^1$, Izumi Watanabe$^1$, Yuki Kawasaki$^1$, Yuta Michimura$^2$, Shuichiro Yokoyama$^{3,4}$, Seiji Kawamura$^{1,3}$}%

\affiliation{%
 1 Department of Physics, Nagoya University, Nagoya, Aichi 464-8602, Japan\\
 2 Department of Physics, The University of Tokyo, Bunkyo, Tokyo 113-0033, Japan\\
 3 The Kobayashi-Masukawa Institute for the Origin of Particles and the Universe, Nagoya University, Nagoya, Aichi 464-8602, Japan\\
 4 Kavli IPMU (WPI), UTIAS, The University of Tokyo, Kashiwa, Chiba 277-8583, Japan
}%




\date{\today}

\begin{abstract}
The Juggled interferometer (JIFO) is an earth-based gravitational wave detector using repeatedly free-falling test masses. With no worries of seismic noise and suspension thermal noise, the JIFO can have much better sensitivity at lower frequencies than the current earth-based gravitational wave detectors. The data readout method of a JIFO could be challenging if one adopts the fringe-locking method. We present a phase reconstruction method in this paper by building up a complex function which has a fringe-independent signal-to-noise ratio. Considering the displacement noise budget of the Einstein Telescope (ET), we show that the juggled test masses significantly improve the sensitivity at 0.1-2.5$\,$Hz even with discontinuous data. The science cases brought with the improved sensitivity would include
detecting quasi-normal modes of black holes with $10^4-10^5\,M_{\odot}$, testing Brans-Dicke theory with black-hole and neutron-star inspirals, and detecting primordial-black-hole-related gravitational waves.

\end{abstract}

\maketitle


\section{\label{sec:1}Introduction}
The current earth-based laser interferometric gravitational wave (GW) detectors keep detecting gravitational waves with best sensitivity at $\sim\hspace{-2mm}100\,\rm{Hz}$~[\onlinecite{aLigo, aVirgo, Kagara, 2020LIGO, GWTC-2}]. With longer arm length and improved techniques, next-generation detectors will have sensitivity down to h$\,\sim\hspace{-0.1mm}10^{-24}/\rm\sqrt{Hz}$~[\onlinecite{ET2010}, \onlinecite{ET2011}]. However, the sensitivity at lower frequencies (f$\,<\,$1$\,$Hz) is still limited by seismic noise and suspension thermal noise. Seismic noise is related to the natural phenomena and human activities. Suspension thermal noise comes from the suspension system used by the current GW detectors to isolate the test masses from the earth. As frequency decreases, both of these two noises increase.

To achieve lower-frequency GW detection, several space-based gravitational wave observatories including LISA~[\onlinecite{LISA}], DECIGO~[\onlinecite{Decigo2001}, \onlinecite{Decigo_b_decigo}], Taiji~[\onlinecite{Taiji2017}, \onlinecite{Taiji2021}], and Tianqin~[\onlinecite{Tianqin2016}] are in progress. These space projects are designed to be sensitive in the frequency range from 1$\,$mHz to 10$\,$Hz. By monitoring the arrival time of signals from millisecond pulsars, the Pulsar Timing Arrays~[\onlinecite{PPTA2015}, \onlinecite{EPTA2015}] can detect GWs ranging from $10^{-9}$-$10^{-7}\,$Hz. By detecting the B-mode polarization in Cosmic Microwave Background~[\onlinecite{CMB1994, CMB1996, CMB1998}], primordial GWs with a wavelength of cosmic scale can be detected indirectly. 

There are also other methods to improve the sensitivity of GW detection at lower frequencies on earth, such as atom interferometry~[\onlinecite{Atomic2008}], suspension point interferometry~[\onlinecite{SusPoint2004}], torsion bar antenna~[\onlinecite{Totsionbar2010}, \onlinecite{Tortionbar2014}], and juggled interferometer (JIFO)~[\onlinecite{JIFO}]. Here we focus on the JIFO, which uses repeatedly free-falling test masses. With this juggling method, the earth-based GW detectors can be free of suspension thermal noise as well as seismic noise, and, therefore improve their sensitivity significantly at lower frequencies. In addition, JIFO can be a good test bed on earth for the space project of GW detection, such as DECIGO (possibly for the quantum noise investigation~[\onlinecite{YAMADA}]), because of the free-falling state of the test masses.

In the previous study~[\onlinecite{JIFO}], the basic setup and data analysis method of the JIFO were investigated. In this paper, more details about the data readout method are provided and the science bonus of a JIFO in GW detection is discussed, assuming implementation of juggled test masses in an Einstein Telescope (ET)~[\onlinecite{ET}]-like interferometer. This paper is organized as follows. In Sec.\ref{sec:2}, we introduce the conceptual design of a JIFO. Then in Sec.\ref{sec:JIFOreadout}, two methods to obtain GW signals with a JIFO are analyzed, followed by the discussion of promising science cases with juggled test masses in Sec.\ref{sec:science}. We finish this paper with a brief conclusion and outlook.

\section{\label{sec:2}Conceptual design of a juggled interferometer}
A JIFO is basically a Michelson-type interferometer with juggled test masses. Like all the other laser interferometric gravitational wave detectors, GW signal is detected by the JIFO through the interference of two laser beams. However, a JIFO does not have Fabry-Perot cavities in its two arms. At lower frequencies, mirror displacement noise dominates the noise budget of an earth-based gravitational wave detector. Since the application of Fabry-Perot cavities accumulates the GW-induced cavity length change as well as the mirror displacement noise, the SNR is not improved with the Fabry-Perot cavities for the low-frequency band.

\begin{figure}[h]
\includegraphics[width=6cm]{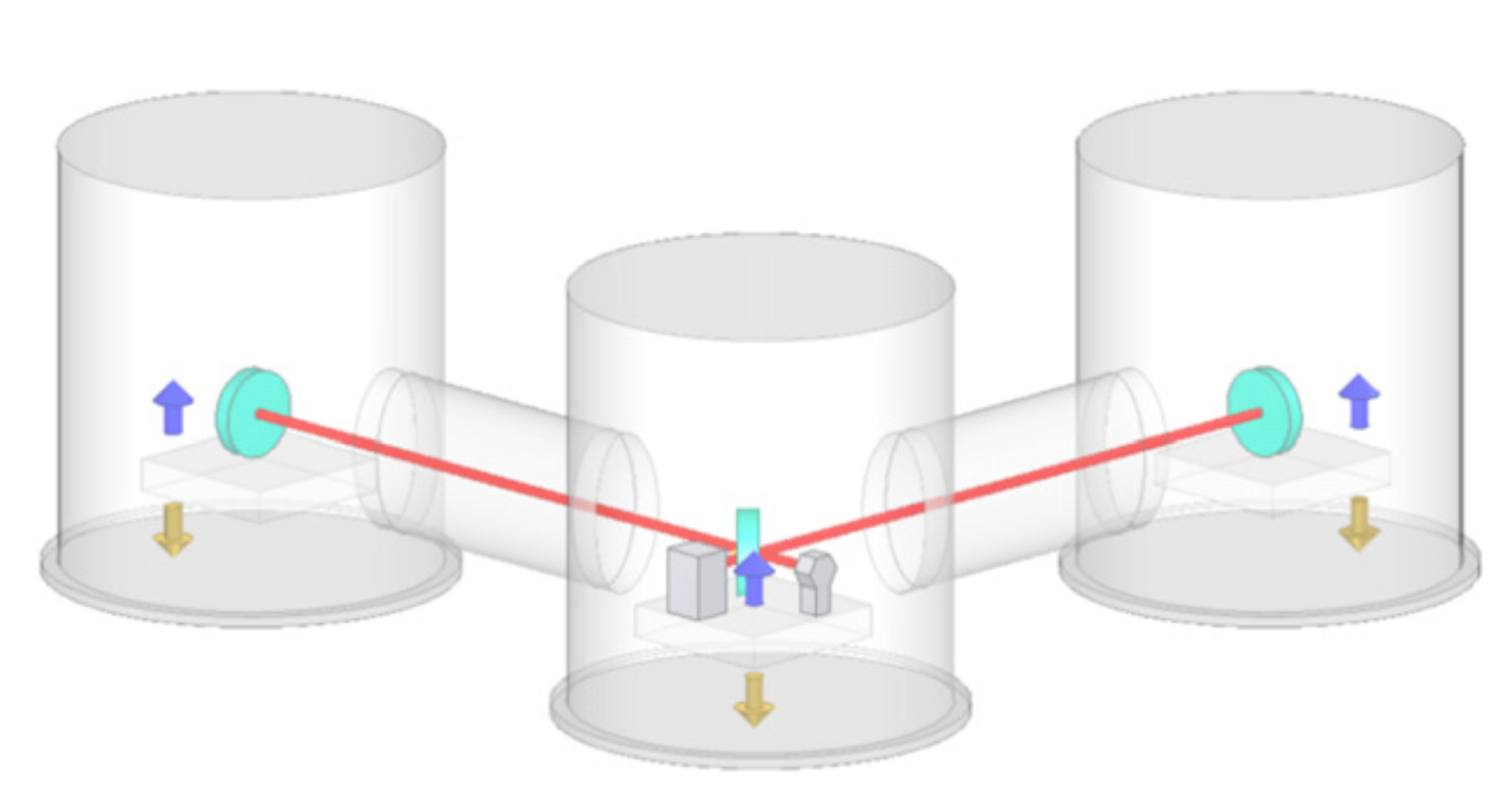}
\caption{\label{fig:conceptual}Conceptual design of a Juggled Interferometer~[\onlinecite{JIFO}].}
\end{figure}

The basic setup for a JIFO is shown in Fig.\ref{fig:conceptual}. The optics are attached to a linear motion guide with clamps and experience repetitive moving processes. Firstly the optics are accelerated upwards along with the slider on the guide . After reaching a certain velocity, the clamps are loosened and the optics are released to proceed a free-falling cycle. The clamps fasten again when the optics are back, and the slider will move again for the next cycle. In this repeatedly accelerated and released operation, the effective data will be discontinuous because only the data collected during the free-falling cycle is free of seismic noise.

\begin{figure}[h]
\includegraphics[width=8cm]{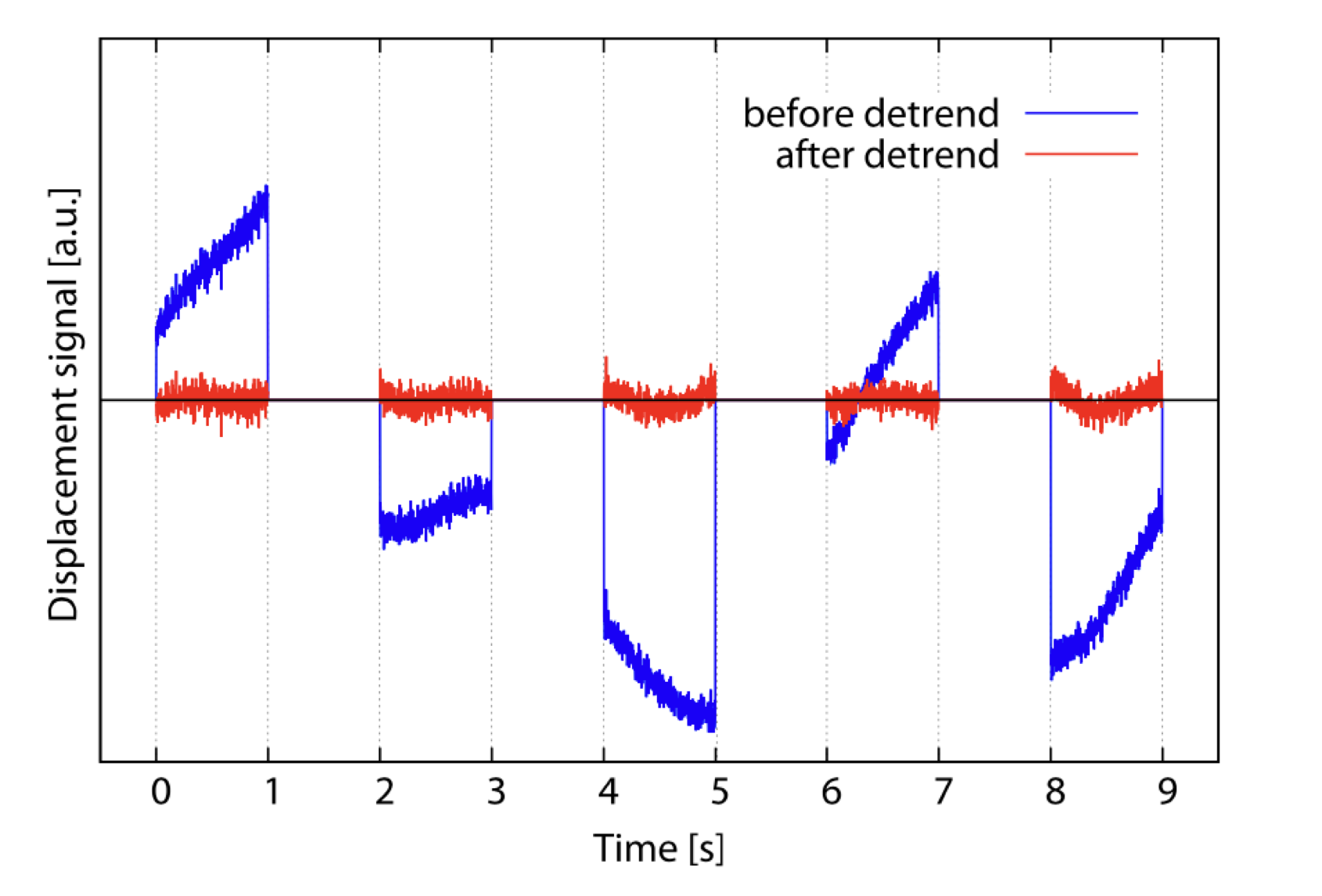}
\caption{\label{fig:detrend}Displacement signal obtained by a JIFO before (blue) and after (red) the detrend process. The intervals between the data indicate the non-data periods during the acceleration cycles~[\onlinecite{JIFO}].}
\end{figure}

Please note that the initial velocity and position of the test masses in each release could differ. This information will exist in the displacement data together with GW signals. So we will first undergo a ``detrend'' method developed in the previous research~[\onlinecite{JIFO}]. The blue curves in Fig.\ref{fig:detrend} are the simulated displacement signal of a JIFO, and the linear trend of the curves is caused by the initial position and velocity of the test masses. By subtracting a linear fit from these blue curves, one can obtain the detrended red curves independent of the initial effects from the test masses.

Inevitably, the GW signals will also be detrended, but this will not influence the signal-to-noise ratio (SNR) for the case of oscillatory signals, since the noises will also be detrended equally. This leads to no requirement for the free-falling time or height. However, for a non-oscillatory signal like the gravitational memory effect~[\onlinecite{gme}], the detrend method is no longer applicable.

The acceleration time is set to be the same as the free-falling time (1$\,$s, for example) in Fig.\ref{fig:detrend}. In principle, acceleration time can be shortened as much as possible. For the case of detecting continuous GWs and considering matched filtering method, if the free-falling time occupies $\frac{1}{N}$ of the total time, then the resulted SNR will be $\frac{1}{\sqrt{N}}$ of that from continuous observation with signal reduced by a factor of $N$ and noise reduced by a factor of $\sqrt{N}$.
By setting up another JIFO with staggered acceleration time, the data intervals can even be filled, making no SNR loss for continuous GWs and greater opportunities for detecting bursts.

\section{\label{sec:JIFOreadout} Interferometer readout and displacement signal reconstruction}

\begin{figure}[h]
\includegraphics[width=8cm]{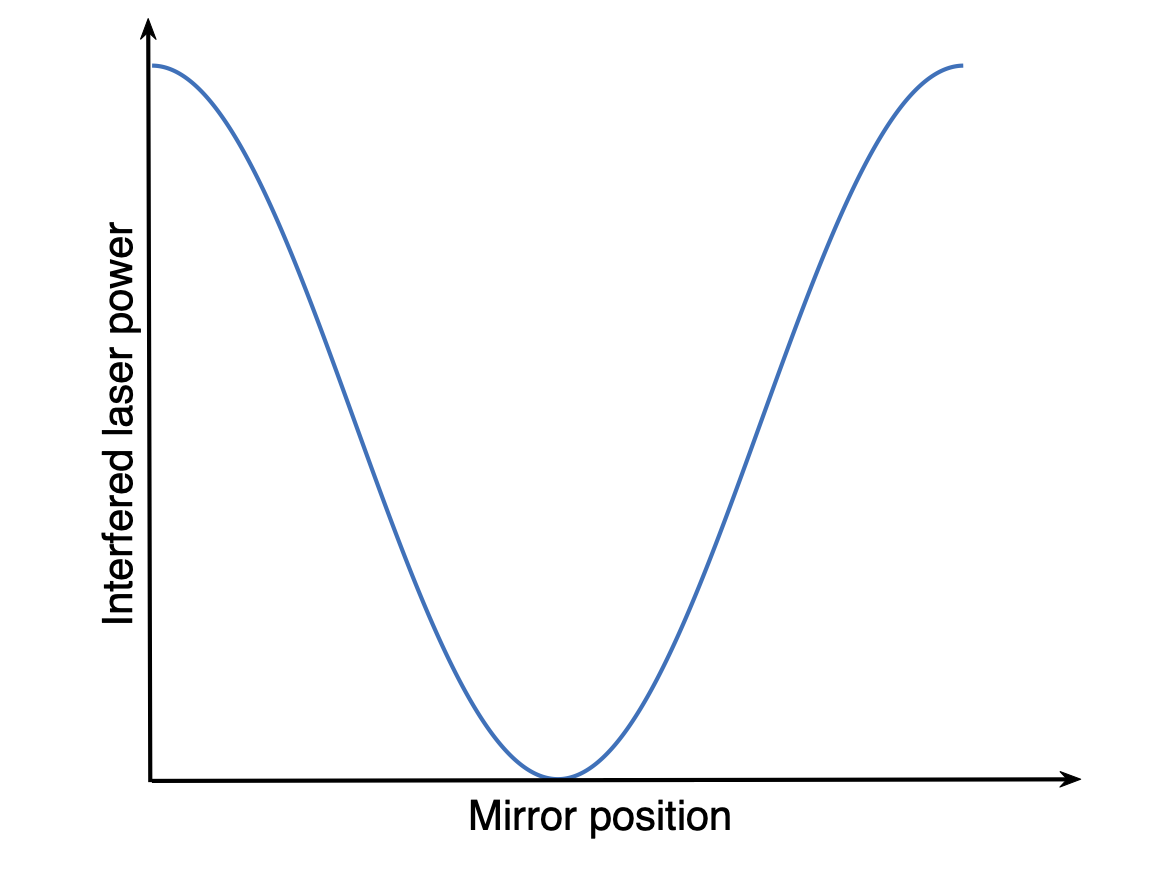}
\caption{\label{fig:posipower}Dependence of the interfered laser power on the test mass position in a cosine wave.}
\end{figure}

In the laser interferometer experiment, the directly measured data is the interfered laser power, while we are more interested in the test mass displacement signal, which straightforwardly indicates the GW signal. Since the interfered laser power varies with the test mass position in a cosine wave (Fig.\ref{fig:posipower}), one can not directly tell the mirror motion from the power variation measured by photodetector (PD). Normally, the interferometric GW detectors use the fringe-locking method to solve this problem. This method is also applied to JIFO but is more challenging. In this section, we discuss the fringe locking of the JIFO and also provide a new reconstruction method to obtain a test mass displacement signal in the JIFO experiment.

\subsection{\label{sec:FringeLocking}Fringe locking of JIFO}
Fringe locking is achieved by controlling the test masses of the interferometer.
For example, if the interferometer is locked at the middle fringe, where the interfered laser power changes linearly to the mirror position, the mirror control signal then indicates the information about the mirror motion. The same thing happens to the control signal when locking at a dark fringe with Pound-Drever-Hall method~[\onlinecite{PDH}]. For the juggled test masses of a JIFO, a remote non-contacting actuator such as an electrostatic drive~[\onlinecite{GEO2003}, \onlinecite{LIGO-ESD2011}], would be a good controller. Given the limited free-falling time, fast response of the controller is demanded.

\subsection{\label{sec:Reconstruct}Reconstruct displacement signal from fringe data}
\begin{figure}[h]
\includegraphics[width=7cm]{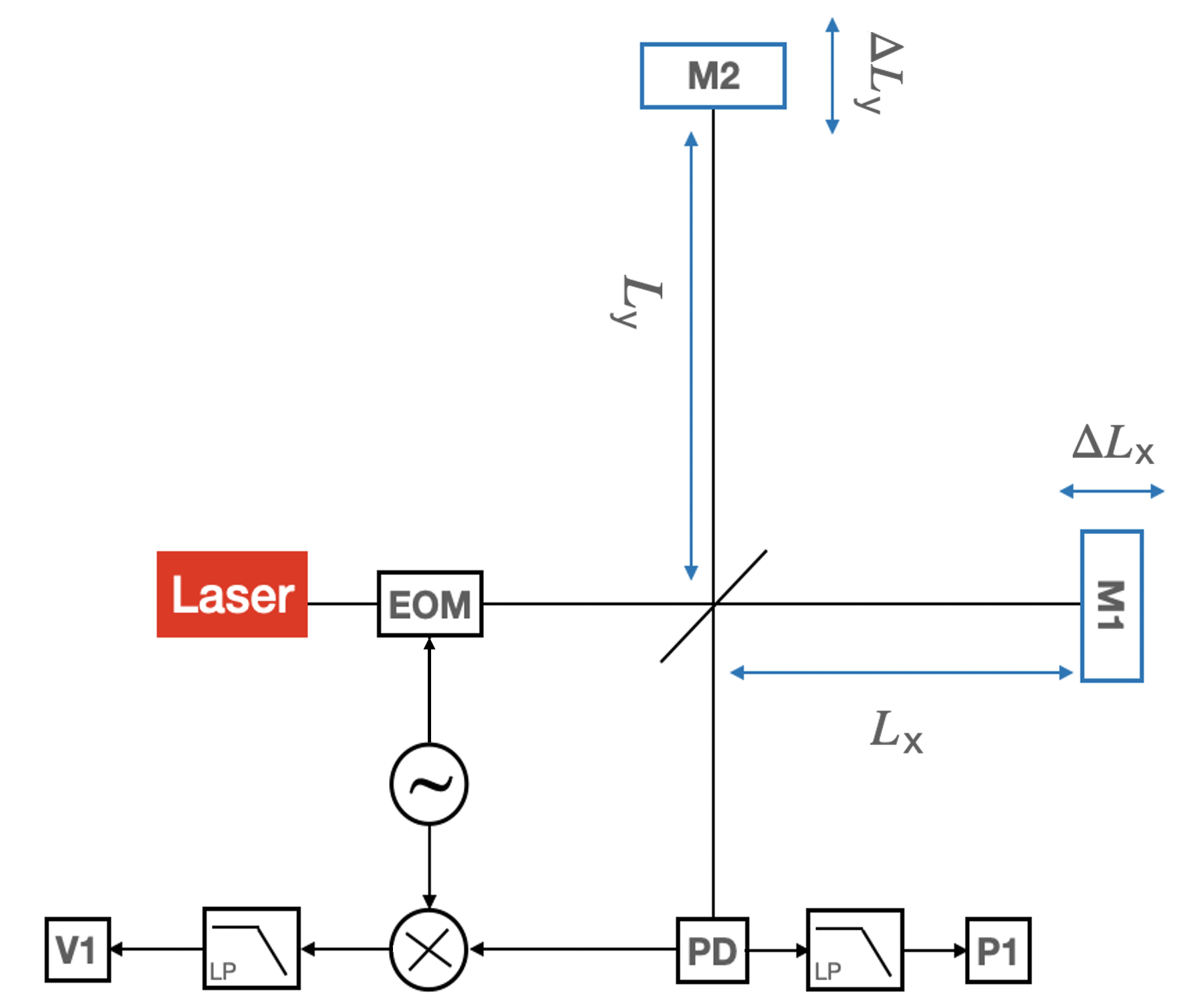}
\caption{\label{fig:setup1}Modulation-demodulation of a laser beam. $P_1$ is the interfered laser power and $V_1$ is the modulation-demodulation signal.}
\end{figure}

The other option is to reconstruct the displacement signal from the laser power signal and modulation-demodulation signal. For the simplicity of the experiment, we avoid juggling the electro-optic modulator (EOM) and, instead, pre-modulate the laser beam before the beam splitter, as is shown in Fig.\ref{fig:setup1}. To maintain mirror motion signal for this pre-modulated interferometer, macro differential arm length $(L_{\rm{x}}\neq L_{\rm{y}})$ is adopted here~[\onlinecite{Schnupp}]. Here we assume the signals are at dark fringe when $t=0 s$. Then the interfered laser power ($P_1$) and modulation-demodulation signal ($V_1$), normalized in power will be:
\begin{eqnarray}
P_1\approx&&\frac{P_0}{2}(1-\rm{cos}(\mathit{\phi(t)})\label{eq:P},
\\
V_1\approx&&-\frac{P_0m_{\rm{eff}}}{2}\rm{sin}(\mathit{\phi(t)})%
\label{eq:V},
\end{eqnarray}
where $P_0$ is the input laser power and $m_{\rm{eff}}=m\sin(\Phi_0)$ with $m$ the modulation depth. $\Phi_0=\left| L_{\rm{y}}-L_{\rm{x}} \right|\times\frac{2\pi}{\lambda_m}$ is the macro phase difference caused by differential arm length with $L_{\rm{x}}$ and $L_{\rm{y}}$ representing the initial arm length in the x direction and y direction. $\lambda_m$ is the wavelength of the modulation signal. $\phi(t)=2 \left| \Delta L_{\rm{y}}-\Delta L_{\rm{x}} \right|\times\frac{2\pi}{\lambda}$ is the phase change introduced by the mirror motion which is the signal that we want to reconstruct. Signals at frequencies higher than the modulation frequency are filtered.  

For the convenience of the calculation, here we ignore the constants in Eq.\ref{eq:P} and Eq.\ref{eq:V}, and adopt only the cosine or sine part:
\begin{eqnarray}
p_1(t)=\rm{cos}\mathit{(\phi(t))}\label{eq:pp},
\\
v_1(t)=\rm{sin}(\mathit{\phi(t)})
\label{eq:vv}.
\end{eqnarray}
The phase signal $\phi(t)$ can be obtained by 
\begin{equation}
\phi(t)={\rm{tan^{-1}}}(\frac{v_1}{p_1})+m\pi+\phi_0,\,\,\,m=\,1,2,3 ...
\end{equation}
and the integer $m$ is chosen with a phase unwrapping algorithm to ensure a continuous phase change~[\onlinecite{phase_unwrap}]. However, this method is complicated when the mirror, and therefore the phase signal, is oscillated with time. This issue can be avoided  by building two complex functions with the interfered laser power (Eq.\ref{eq:pp}) being the real part and the modulation-demodulation signal (Eq.\ref{eq:vv}) being the imaginary part:
\begin{eqnarray}
p_1+iv_1=e^{i\phi(t)}\label{eq:eit},
\\
p_1-iv_1=e^{-i\phi(t)}
\label{eq:e-it},
\end{eqnarray}
thus the differential of $\phi(t)$ can be obtained from: 
\begin{equation}
    \frac{d\phi(t)}{dt}=-i\times\frac{de^{i\phi(t)}}{dt}\times e^{-i\phi(t)},\label{eq:diffphi}
\end{equation}
while $\phi(t)$ can be reconstructed by integration the equation of over time:
\begin{equation}
    \int _0^t\frac{d\phi(t)}{dt}dt=\phi(t)+\rm{const.}\label{eq:intephi}
\end{equation}
Note that we can not determine the initial phase difference with this method which indicates the initial position of the test mass. Also note that high frequency noise would be introduced from the differentiation in Eq.\ref{eq:diffphi} but can be removed by the integration in Eq.\ref{eq:intephi}.

\begin{figure}[h]
\includegraphics[width=7cm]{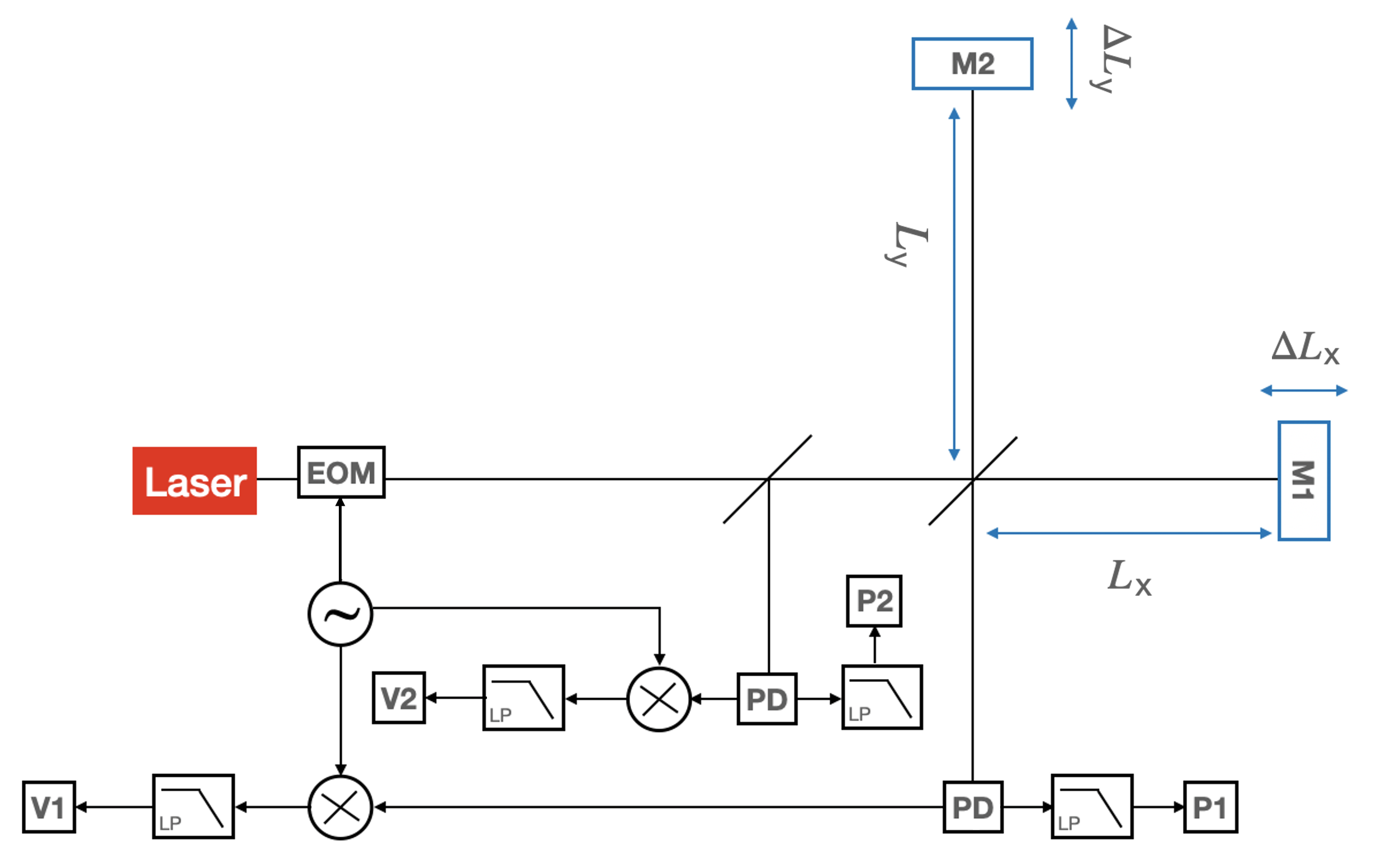}
\caption{\label{fig:setup2}Extension of Fig.\ref{fig:setup1} by collecting light power ($P_2$) and modulation-demodulation signal ($V_2$) on the laser-side port.}
\end{figure}

It is worth discussing the SNR of the data readout with the complex function when we take the modulation-demodulation signal from the laser-side port ($V_2$ in Fig.\ref{fig:setup2}). In this case, we have $v_2=-v_1$, and
\begin{equation}
    e^{i\phi(t)}=p_1+iv_1=p_1-iv_2.
\end{equation}
The shot noise in terms of the laser power is usually calculated by:
\begin{equation}
    n_{shot} = h\nu\times\tilde{N}=\sqrt{\frac{Phc}{\lambda}},
\end{equation}
where $\tilde{N}=\sqrt{\frac{P\lambda}{hc}}[1/s]$ is the photon number fluctuation, $h$ is the Plank constant, $c$ is the speed of light, $P$ is the laser power and $\nu$ is the laser frequency. Since the shot noise from the two PD ports are uncorrelated, the total shot noise in terms of the laser power change can then be calculated by: 
\begin{equation}
   \sqrt{\sqrt{\frac{P_1hc }{\lambda}}^2+\sqrt{\frac{P_2hc }{\lambda}}^2}=\sqrt{\frac{P_0hc}{\lambda}} .
\end{equation}
 We can see that in this case, the shot noise is determined by the input laser power $P_0$ and does not change with the phase signal. Considering only shot noise, the SNR of ($p_1-iv_2$) will be:
\begin{eqnarray}
&&{\rm{SNR}}(e^{i\phi(t)}=p_1-iv_2)\nonumber
\\=&&\frac{P_0/2\times \left| de^{i\phi(t)}\right|}{\sqrt{\frac{P_0hc}{\lambda}}}\nonumber\\
=&&\frac{P_0|d\phi(t)|\times	\left| {\rm cos}(\phi(t))+i{\rm sin}(\phi(t))\right| }{2\sqrt{\frac{P_0hc}{\lambda}}}\nonumber\\
=&&\frac{|d\phi(t)|}{2}\sqrt{\frac{P_0\lambda}{hc}}.
\end{eqnarray}
The same result can be derived by the combination of $p_2$ and $v_1$. Here we assume the shot noise of the modulation-demodulation signal on the laser-side port ($V_2$) is consistent with that of the interfered power laser ($P_2$) and uncorrelated with $P_1$. It is interesting to see that the SNR of this combination is fringe-independent and is half of the SNR at the dark fringe. 

Compared with the fringe-locking method mentioned in Sec.\ref{sec:FringeLocking}, the reconstruction method is much more convenient since there is no need to lock the test masses anymore. But we have to pay attention to the ADC (Analog to Digital Converter) resolution with the demand of large output voltage scale ranging from the dark fringe to the bright fringe. In addition, the output laser power would be too large for a single PD and may need to be shared by multiple PDs.

\section{\label{sec:science}Promising science cases of a juggled interferometer}
\begin{figure}[h]
\includegraphics[width=8cm]{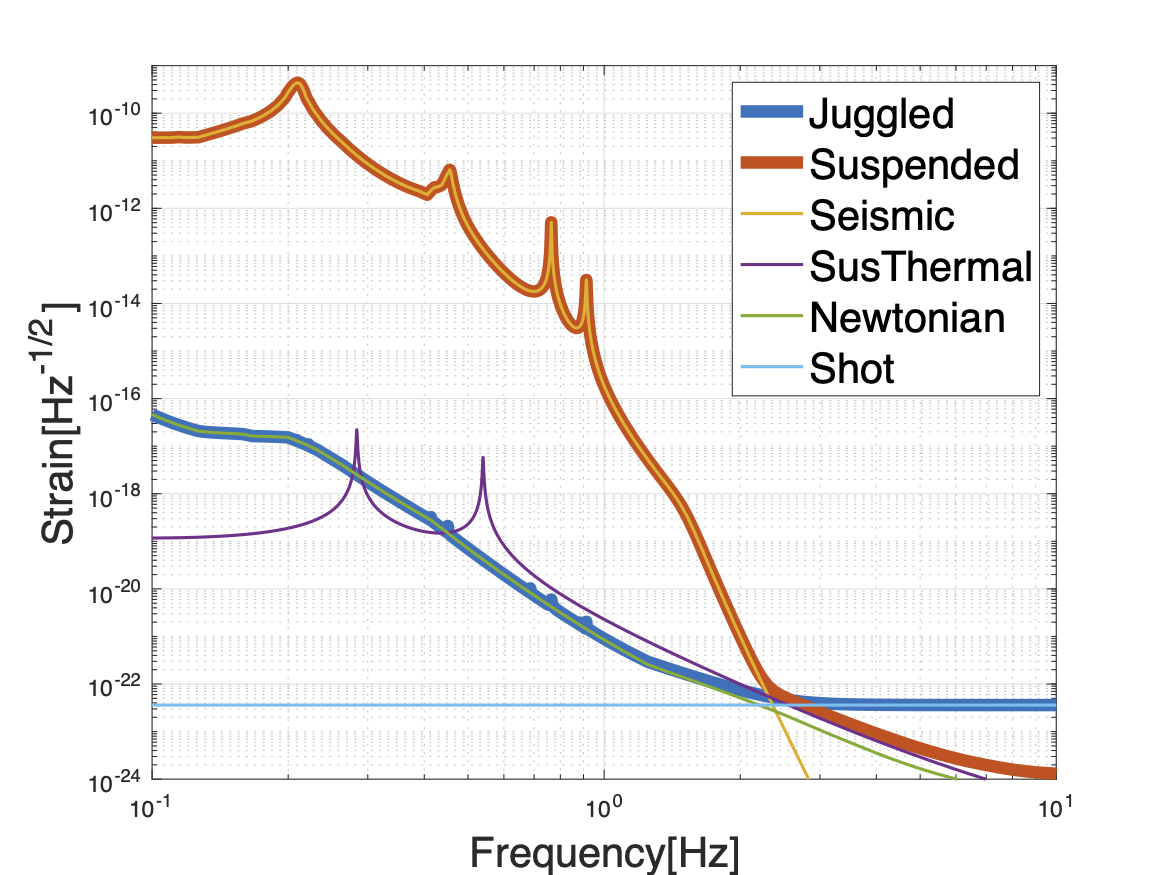}
\caption{\label{fig:NB}Noise budget of the JIFO based on ET's displacement noise. The bold blue curve shows the budget of JIFO (free-falling/juggled test masses), while the bold red curve shows the budget of ET (suspended test masses). Without seismic noise (orange) and suspension thermal noise (purple), the juggled case lowers the noise budgets significantly below $\sim$2.5$\,$Hz. The dominant noise of JIFO at lower frequencies is Newtonian noise (green), and shot noise (cyan) limits the sensitivity at higher frequencies.}
\end{figure}

In this section, we assume a Michelson-type interferometer with the  same scale and displacement-noise budget as the ET~[\onlinecite{ET2010}, \onlinecite{ET2011}], which is the next-generation gravitational wave detector planned to be operational in the 2030s. The implement of juggled test masses, which means no worries of seismic noise and suspension thermal noise, will result in significant improvement in the detector's sensitivity below $\sim$2.5$\,$Hz, as is shown in Fig.\ref{fig:NB}.
To achieve this sensitivity level, the laser frequency noise should be below $3\times10^{-5} \rm Hz/\sqrt{Hz}$ and the laser power fluctuation should be below  $10^{-6}/\sqrt{\rm Hz}$, assuming a length difference of two arms to be within 1 m and the offset from the middle fringe to be $10^{-13}$ m.

 The following discussions will mention several promising science cases for the JIFO with the improved sensitivity below $\sim$2.5$\,$Hz. 

\subsection{\label{sec:QNM}Detecting quasi-normal modes of massive black holes}
The perturbations of black holes, which normally happen when particles fall into them or in the late process of two massive bodies coalescing into a black hole, will generate gravitational waves called black-hole quasi-normal modes (BH QNMs)~[\onlinecite{QNM1}, \onlinecite{QNM5}]. The strain of BH QNMs has an exponential decrease feature indicating that GWs take energy away from the system. 

\begin{figure}[h]
\includegraphics[width=8cm]{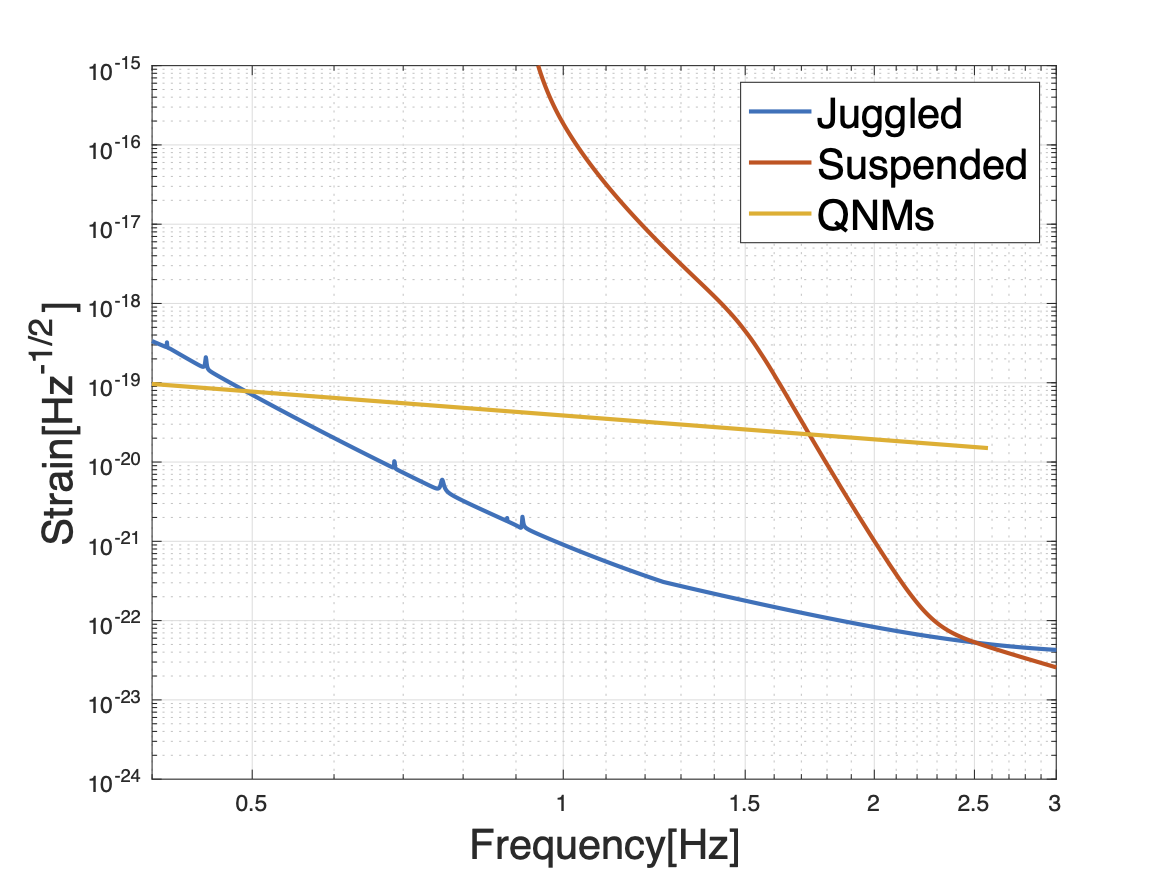}
\caption{\label{fig:QNM}QNMs strain for massive black holes (yellow line) assuming the sources at a distance of 6\,Gpc with the spin a=0.98. The sensitivity curve of JIFO (blue) can achieve QNMs detection in the 0.5-2.5$\,$Hz band with SNR up to 200, while the interferometer with suspended test masses (ET) can not reach them in this frequency band.}
\end{figure}

According to general relativity, the central frequency of a BH QNM is related to the mass of the black hole $M$ and its spin angular momentum $a= [0, 1)$~[\onlinecite{QNM}]:
\begin{equation}
f_{\rm{c}}(\rm Hz)=\frac{32\times 10^3 M_{\odot}}{\mathit{M}}\times[1-0.63(1-\mathit{a})^{0.3}],
\label{eq:QNM}
\end{equation}
and the 0.1-2.5$\,$Hz frequency band corresponds to black holes with $10^4-10^5\,M_{\odot}$. Under the assumption that the signal is monochromatic, the effective dimensionless strain of the QNMs are estimated by~[\onlinecite{QNM2}, \onlinecite{lalsuite}] 
\begin{equation}
h_{\rm eff}\sim 3\times10^{-20}[\frac{E}{250M_{\odot}c^2}]^{1/2}[\frac{f_{\rm{c}}}{1{\rm Hz}}]^{-1/2}[\frac{r}{6{\rm Gpc}}]^{-1},
\label{eq:QNM-h}
\end{equation}
where $E$ is the emitted energy which is assumed to be 1\% of the mass of the black hole in this paper and $r$ is the luminosity distance. Fig.\ref{fig:QNM} shows the effective strain divided by the square root of frequency, resulting in the same unit with the noise budget. The BH QNMs in 0.5-2.5$\,$Hz is lying above the improved sensitivity (blue curve) with an SNR up to 200 around 1.7 $\,$Hz according to the matched filtering method, corresponding to the black holes with a mass of about $1.5\times 10^4\,{\rm{M_{\odot}}}$. These sources are beyond the detection ability before the improvement (red curve).

BH QNMs have been extensively discussed as a powerful tool to test the theory of gravity~[\onlinecite{QNM4}, \onlinecite{QNM5}]. The comparison between the observed signals and the predictions can be a convincing test of different gravitational theories.

\subsection{\label{sec:BD}Testing Brans-Dicke theory}

\begin{figure}[h]
\includegraphics[width=8cm]{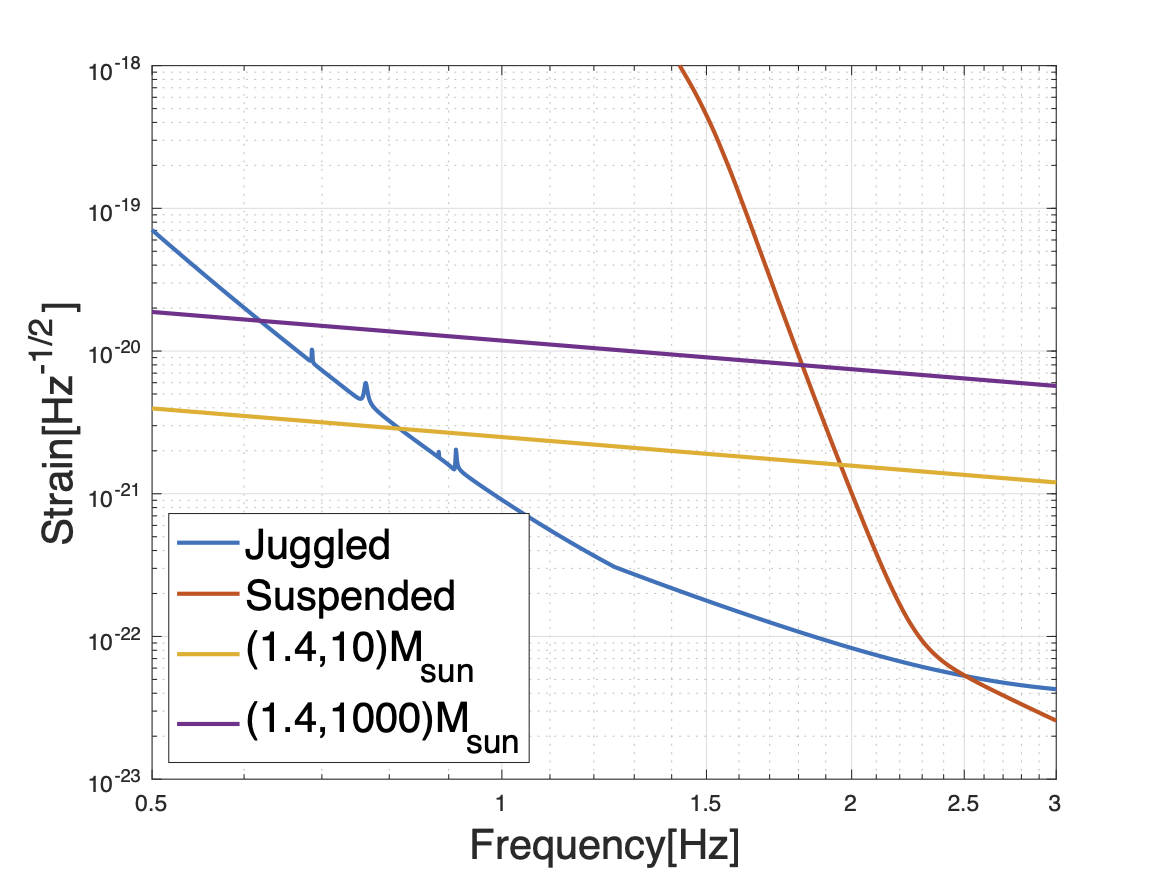}
\caption{\label{fig:BD}Gravitational wave strain from the inspiral of NS-BH binaries at a distance of 78$\,$Mpc.}
\end{figure}

In Brans-Dicke (BD) theory~[\onlinecite{BD2}], the scalar field and tensor field both mediate the gravitational interaction. When the BD parameter $\omega_{\rm{BD}}$ approaches infinite, BD theory is back to general relativity. A GW signal from the inspiral of two dense objects, with one of which is a neutron star, is used to constrain $\omega_{\rm{BD}}$~[\onlinecite{BD}]. Fig.\ref{fig:BD} shows that the improved sensitivity curve (blue) has more chances to detect NS-BH (neutron star and black hole) inspirals below 2.5$\,$Hz. 

Following the method in Refs.~[\onlinecite{BD},~\onlinecite{ BD-2022}], we obtained the lower bound of $\omega_{\rm{BD}}$ as $8.8\times 10^4$ by assuming the detection of a $(1.4,10)M_{\odot}$ NS-BH binary below 2.5 Hz at 78 Mpc which gives SNR=10, according to the matched filtering method. See the appendix \ref{appendix-BD} for more details. This event rate is believed to be 0.01-1 in 5 years~[\onlinecite{eventrate}]. As a comparison, the solar system experiment using the Saturn probe satellite Cassini obtained  $\omega_{\rm{BD}}>4\times10^4$~[\onlinecite{cassini}] and the space GW detector DECIGO could obtain $\omega_{\rm{BD}}>2\times10^6$~[\onlinecite{BD}]. Considering multiple events with frequency over 1 Hz, the bound from ET could be $\omega_{BD}>10^6$[\onlinecite{BD_ET}].

\subsection{\label{sec:PBH}Primordial black holes research}

Primordial black holes (PBHs), which would be directly formed from the
primordial curvature perturbations in the radiation-dominated universe~[\onlinecite{Hawking}, \onlinecite{Carr}], have been attracting attention in
recent years. Unlike black holes originating from astrophysical
processes, which must be massive enough, PBHs could have been formed to
have a vast mass range, and quite a number of PBHs lighter than the Sun could
exist. In particular, such a PBH with the mass of an asteroid
(\(\sim 10^{17}-10^{22}~{\rm g}\)) is extensively discussed as a
candidate of the dominant component of the dark matter in our universe
(see, e.g., [\onlinecite{Carr_2021}]). The primordial curvature perturbations
as a seed for the PBH formation can also generate the stochastic
gravitational wave background (denoted as PBH-GW)~[\onlinecite{PBH1}],
and the central frequency of such a PBH-GW is related to the mass of
formed PBH as

\begin{equation}
f_{\rm GW} = 1~{\rm Hz}\, \left( \frac{M_{\rm PBH}}{10^{17}\,{\rm g}} \right)^{-1/2}
\left( \frac{g_{\ast}}{106.75} \right)^{-1/12},
\label{eq:PBH1}
\end{equation}
in which \(M_{\rm PBH}\) is the mass of PBH and \(g_{\ast}\) is the relativistic
degree of freedom at the time of PBH formation. Thus, the observation of the gravitational wave background in the \(\sim 1\) Hz band is very
important as an indirect probe of primordial black holes as dark matter.

PBH would be basically formed at a high and rare peak of the primordial perturbations and then the abundance of PBHs is roughly proportional to \(\exp[-\zeta_{\rm th}^2/A_\zeta^2]\) in which \(\zeta_{\rm th}\)
represents the threshold for the PBH formation and the \(A_\zeta\) is
the typical amplitude of the primordial curvature perturbations. The
value of the threshold, \(\zeta_{\rm th}\), is typically ${O}(0.1)$, and in order for PBH to be dark matter the amplitude \(A_\zeta^2\) needs
to be roughly \(\sim 10^{-2}-10^{-3}\) (see, e.g.,~[\onlinecite{Yoo}]). On the other hand, the PBH-GW is sourced from the
second order of the primordial curvature perturbations and the density
parameter of GWB, \(\Omega_{\rm GW}\), is proportional to \(A_\zeta^4\).

\begin{figure}[h]
\includegraphics[width=8cm]{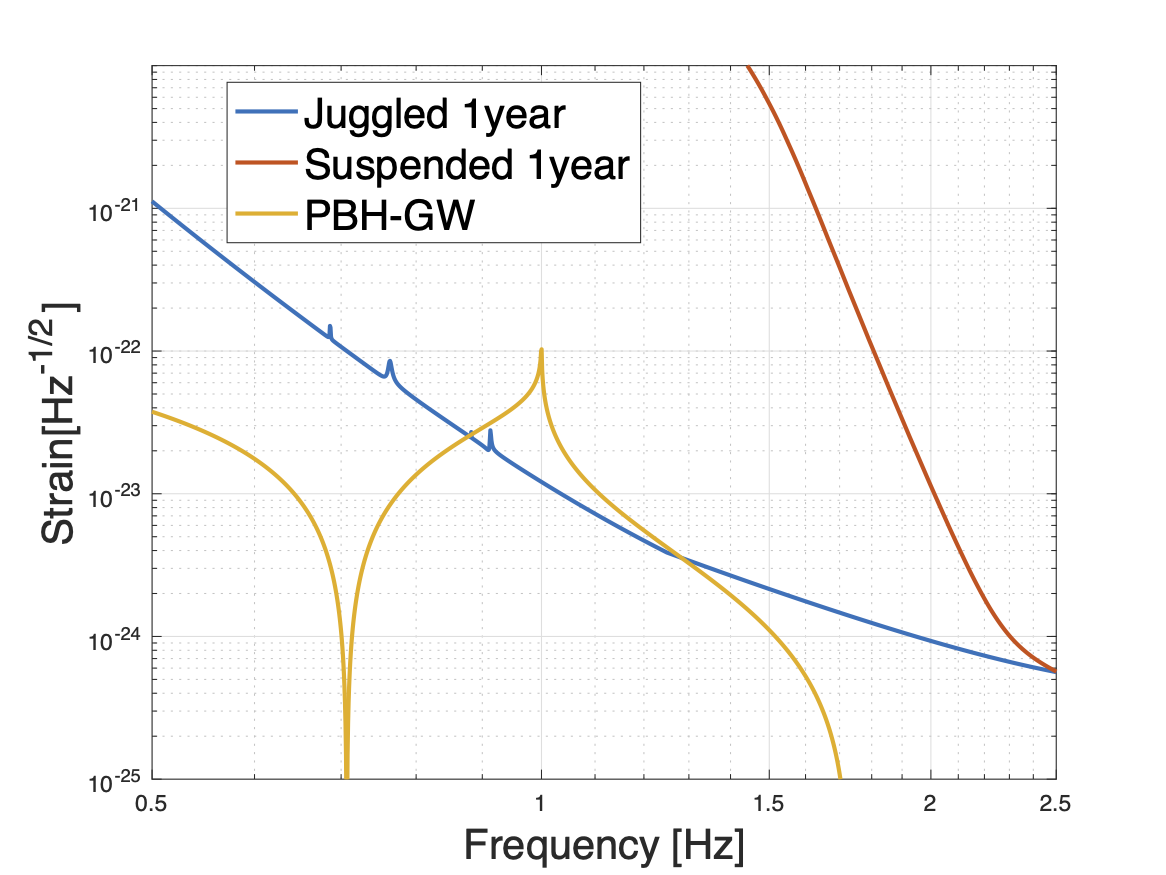}
\caption{\label{fig:PBH}Strain of a PBH-GW peaked at 1$\,$Hz with $A_{\zeta}^2=5\times10^{-3}$ and the detector sensitivity with 1 year correlation. The primordial density fluctuation is assumed to be a delta function.}
\end{figure}

 Here, for simplicity, following Refs. [\onlinecite{PBH1}, \onlinecite{{Kohri}}], the energy density of the PBH-GW is calculated by
\begin{equation}
\begin{aligned}
\Omega_{\mathrm{GW}}(\eta, k)=& \frac{3 A_{\zeta}^{2}}{64}\left(\frac{4-\tilde{k}^{2}}{4}\right)^{2} \tilde{k}^{2}\left(3 \tilde{k}^{2}-2\right)^{2} \\
& \times\left(\pi^{2}\left(3 \tilde{k}^{2}-2\right)^{2} \Theta(2 \sqrt{3}-3 \tilde{k})\right.\\&\left. +\left(4+\left(3 \tilde{k}^{2}-2\right) \log \left|1-\frac{4}{3 \tilde{k}^{2}}\right|\right)^{2}\right) \Theta(2-\tilde{k}),
\end{aligned}
\label{PBH:energy density}
\end{equation}
under the assumption that the primordial curvature perturbations have the monochromatic power spectrum with a delta function peak at the wavenumber of $k_*$. The dimensionless wavenumber $\tilde{k}\equiv k/k_*$, where $k$ is the wavenumber of the gravitational wave signal.
Then the amplitude spectrum density (ASD) of the PBH-GW, $S_h(f)$, can be derived by [\onlinecite{asd}]:
\begin{equation}
    S_h(f) = \Omega_{\mathrm{GW}}\frac{3H_0^2}{2\pi^2f^3},
\end{equation}
where $H_0$ is the Hubble expansion rate. Fig.\ref{fig:PBH} shows the ASD of a PBH-GW peaked at \(1\) Hz with the amplitude $A_\zeta^2 = 5\times10^{-3}$. From this figure, the PBH-GW is detectable by the JIFO with a time correlation of \(1\) year and the detection SNR reaches 19.
Note that this result relies on the shape of the spectrum of PBH-GW and strongly depends on the assumption for the primordial curvature
perturbations. Nevertheless, the JIFO, the implement of free-falling
test masses into an ET-like interferometer, will
provide us with important insights into cosmology.

\section{\label{sec:sum}Summary and outlook}
To improve the sensitivity of the earth-based GW detectors, JIFO removes the seismic noise and suspension thermal noise with the juggled test masses. There are two methods to obtain mirror motion signal from a JIFO. One is the fringe-locking method, and it obtains the mirror motion information from the control signal. The other method is to modulate and demodulate the laser beam and reconstruct the mirror motion information from interfered power and modulation-demodulation signal. The SNR of the latter is proved to be fringe-independent and is half of that at a dark fringe. 

Based on the noise budget of ET, the implementation of juggled test masses will improve the sensitivity significantly below 2.5$\,$Hz. A short free-falling distance would be enough for the desired noise level. This sensitivity improvement will provide promising detection cases such as QNMs from massive black holes, GWs from NS-BH inspirals, and GWs related to PBH-DMs. These detections will provide a new method for testing testing the gravitational theories and cosmology.

There are still some technical difficulties, though. First, the free-falling system plays high demands on the accuracy of mirror control. Second, locking the juggled test masses will be challenging if we decide to adopt the fringe-locking method, while the ACD resolution could be a problem if we adopt the modulation-demodulation method. We plan to build a prototype of JIFO based on the conceptual design in this paper.

\begin{acknowledgments}
We would like to thank Michael E. Zucker for the English editing. This work was supported by the Japan Society for the Promotion of Science (JSPS) KAKENHI (grant number 21K18626), Murata Science Foundation, and the program of China Scholarships Council. S.Y. is supported by JSPS Grant-in-Aid for Scientific Research (C) No. JP20K03968.
\end{acknowledgments}

\appendix
\section{Constrain Brans-Dicke Parameter}\label{appendix-BD}

Here we elaborate on the method we adopted to constrain Brans-Dicke parameter~[\onlinecite{BD},~\onlinecite{ BD-2022}]. 
The Fourier component of the waveform under stationary phase approximation is given by:

\begin{equation}
\tilde{h}(f)=\frac{\sqrt{3}}{2} \mathbb{A} f^{-7 / 6} e^{i \Psi(f)}
,
\label{eq:bd-h}
\end{equation}
with the amplitude
\begin{equation}
\mathbb{A}=\frac{1}{\sqrt{30} \pi^{2 / 3}} \frac{\mathbb{M}^{5 / 6}}{D_{L}}
,
\label{eq:bd-a}
\end{equation}
and the phase expanded to the second post-Newtonian (PN) order:
  \begin{equation}
\begin{aligned}
\Psi(f)=& 2 \pi f t_{c}-\phi_{c}+\frac{3}{128}(\pi \mathbb{M} f)^{-5 / 3}\left[1-\frac{5}{84} \mathbb{S}^{2} \bar{\omega} x^{-1}\right.\\&\left.+\left(\frac{3715}{756}+\frac{55}{9} \eta\right) x-4(4 \pi-\beta) x^{3 / 2}\right.\\&\left.+\left(\frac{15293365}{508032}+\frac{27145}{504} \eta+\frac{3085}{72} \eta^{2}-10 \sigma\right) x^{2}\right],
\end{aligned}
\label{eq:bd-phi}
\end{equation}
where $f$ is the gravitational wave frequency, $\mathbb{M}$ is the chirp mass, $D_L$ is the luminosity distance from the source to the detector, $t_c$ and $\phi_c$ are the time and phase of the coalescence and $\beta$ and $\sigma$ indicates the spin-orbit and spin-spin contributions to the phase respectively.

The other parameters in the waveform are defined as following: $\mathbb{S}\equiv s_2-s_1$ with $s_i$ the $sensitivity$ of the $i_{th}$ body of the binay system which is roughly the binding energy of the body per unit mass,  $\bar{\omega}\equiv1/\omega_{BD}$ is the inverse of the Brans-Dicke parameter, 
$\eta \equiv\frac{m_1m_2}{(m_1+m_2)^2}$ with $m_i$ the mass of the $i_{th}$ body and $x=[\pi(m_1+m_2)f]^{2/3}$. Note that a relatively large $\mathbb{S}$ can be obtained from two different types of bodies. For example, BH-NS system gives $\mathbb{S}\sim 0.3$ while BH-BH/NS-NS system gives $\mathbb{S}\sim 0$. The gravitational wave signal from a BH-NS system is usually used to estimate the Brans-Dicke parameter because the larger $\mathbb{S}$ resulted in the lager contribution form the $\bar{\omega}$ to the waveform.

The standard parameter estimation method in matched filtering is adopted here to constrain the Brans-Dicke parameter. For a waveform defined by a set of parameters $\boldsymbol{\theta}=(\theta_i,\theta_j, ...)$, the Fisher matrix is derived by
\begin{equation}
\Gamma_{i j} \equiv\left(\frac{\partial h}{\partial \theta_{i}} \mid \frac{\partial h}{\partial \theta_{j}}\right).
\label{eq:Fisher-matrix}
\end{equation}
The inner product here is defined as
\begin{equation}
(A \mid B)=4 \operatorname{Re} \int_{0}^{\infty} d f \frac{\tilde{A}^{*}(f) \tilde{B}(f)}{S_{n}(f)},
\label{eq:inner-product}
\end{equation}
where $S_{n}(f)$ is the noise power spectral density of the detector. Thus the diagonal elements of the Fisher matrix indicate the accuracy of the estimation of the corresponding parameters and the rms error of the estimation can be calculated by taking the sqaure root of the diagonal elements of the inverse of the Fisher matrix. Since the expected value of $\bar{\omega}$ is 0, the rms error is then its upper limit. Therefore the lower limit of $\omega_{BD}$ can be obtained from the inverse of the rms error. 

Here we consider the detection of a $(1.4,10)M_{\odot}$ NS-BH binary below 2.5 Hz at 78 Mpc. The sensitivity of the JIFO shown in Fig.~\ref{fig:NB} gives SNR=10 with the assumption of $t_c=0, \phi_c=0, \mathbb{S}=0.3, \bar{\omega}=0, \beta=0, \sigma = 0$. And the diagonal element of the inverse Fisher matrix corresponding to $\bar{\omega}$ is $1.14\times10^5$, showing that the lower bound of $\omega_{BD}$ is $8.8\times10^4$.


\bibliography{jifo_wu}

\end{document}